\documentclass[%
mathleft,%
final,%
]{an}

\usepackage{graphicx}
\usepackage{times}
\usepackage{epstopdf}
\begin{document}
\Pagespan{133}{136}
\Yearpublication{2013}%
\Yearsubmission{2012}%
\Month{1}%
\Volume{334}%
\Issue{1/2}%
\DOI{10.1002/asna.201211741}%

\title{Three-dimensional magnetohydrodynamic simulations of M-dwarf chromospheres}

\author{S. Wedemeyer\inst{1,2}\fnmsep\thanks{Corresponding author:
  \email{sven.wedemeyer@astro.uio.no}}
\and  H.-G. Ludwig\inst{3}
\and  O. Steiner\inst{4}
}
\titlerunning{Three-dimensional magnetohydrodynamic simulations of M-dwarf chromospheres}
\authorrunning{Wedemeyer, Ludwig, \& Steiner}
\institute{
Institute of Theoretical Astrophysics, University of Oslo,
  P.O. Box 1029 Blindern, N-0315 Oslo, Norway
  \and
  Center of Mathematics for Applications (CMA), University of Oslo,
  Box 1053 Blindern, N-0316 Oslo, Norway
  \and
  ZAH, Landessternwarte K{\"o}nigstuhl,
  D-69117~Heidelberg,
  Germany
  \and
  Kiepenheuer Institute for Solar Physics, Sch{\"o}neckstrasse 6-7, D-79104 Freiburg, Germany
  }

\received{2012 Jul 11}
\accepted{2012 Dec 5}
\publonline{2013 Feb 1}
\keywords{stars: atmospheres, chromospheres, magnetic fields; convection, magnetohydrodynamics (MHD), radiative transfer, shock waves.}

\abstract{%
We present first results from three-dimensional 
radiation magnetohydrodynamic simulations of M-type dwarf stars with CO5BOLD.  
The local models include the top of the convection zone, the photosphere, and the chromosphere.  
The results are illustrated for models with an effective temperature of 
3240 K and a gravitational acceleration of log g = 4.5, 
which represent analogues of AD Leo. 
The models have different initial magnetic field strengths and field topologies. 
This first generation of models demonstrates that the atmospheres of 
M-dwarfs are highly dynamic and intermittent. 
Magnetic fields and propagating shock waves produce a 
complicated fine-structure, which is clearly visible in 
synthetic intensity maps in the core of the Ca II K 
spectral line and also at millimeter wavelengths. 
The dynamic small-scale pattern cannot be described by means of 
one-dimensional models, which has important implications for the 
construction of semi-empirical model atmospheres and thus for the 
interpretation of observations in general. 
Detailed three-dimensional numerical simulations are valuable in this 
respect. 
Furthermore, such models facilitate the analysis of small-scale processes, 
which cannot be observed on stars but nevertheless might be essential 
for understanding M-dwarf atmospheres and their activity. 
An example are so-called ``magnetic tornadoes'', which have recently been found 
on the Sun and are presented here in M-dwarf models for the first time. 
}

\maketitle

\section{Introduction}

Our picture of the atmospheric fine-structure of our Sun changed substantially  
during the past decades. 
Progress in instrumental performance resulted in unprecedented \linebreak 
high-resolution observations that 
reveal a high degree of structure on a multitude of spatial and temporal scales. 
Such a complex phenomenon cannot be described in sufficient detail by means 
of otherwise elaborate semi-empirical one-dimensional model atmospheres  
(e.g. Vernazza et al. 1981). 
%
The importance of small-scale spatial and temporal variations in the solar 
chromosphere has been realized  
(cf. Ayres \& Rabin 1996; Carlsson \& Stein 1995; Solanki et al. 1991)
and is now an essential feature of state-of-the-art numerical models. 
Furthermore, the layers of the Sun are no longer treated individually but 
as integral components of a dynamically coupled atmosphere 
(Wedemeyer-B\"ohm et al. 2009).

Here we present a new class of 3-D models for M-dwarf stars that 
clearly demonstrates that their atmospheres are dynamic and intermittent, too. 
This finding immediately gives rise to the question to what extent such inhomogeneities 
affect the emergent radiative intensity and related chromospheric diagnostics. 
In order to address this question, we use model snapshots as input for spectrum 
synthesis calculations and discuss the  resulting implications of the spatial and 
temporal inhomogeneities for the interpretation of observations and for the 
construction and usage of \linebreak one-dimensional static model atmospheres.

\section{Numerical simulations}

\paragraph{The code.}
The numerical simulations are carried out with the 
3-D radiation magnetohydrodynamics code \mbox{CO$^5$BOLD} 
(Freytag et al. 2012).
The code numerically solves the equations of (ideal) 
magnetohydrodynamics and  radiative transfer together with 
a realistic equation of state and multi-group opacities. 
The simulation is advanced in time step by step and outputs a 
time sequence of 3-D snapshots and auxiliary data.   
\mbox{CO$^5$BOLD} is applied to a large range of different stars, 
among them the Sun 
(Steiner et al. 2008; Wedemeyer et al. 2004) 
and red giants 
(Freytag et al. 2002). 
%
Wende et al. (2009)
used \mbox{CO$^5$BOLD} to calculate 
local hydrodynamic models of M-type dwarf stars, which all include 
the top of the convection zone and the photosphere 
for a few selected parameter sets of $T_\mathrm{eff}$ and $\log g$ 
(see also 
Beeck et. al 2011). 
The local simulations presented here include in addition a chromosphere 
and magnetic fields. 
The opacity tables are based on  \mbox{PHOENIX} NextGen data 
(Hauschildt et al. 1999).
A particular challenge is that the typical computational time step for M-dwarf 
chromosphere models is on the order of only 1\,ms, owing to the short time scales 
in the magnetized chromosphere. 

\paragraph{The models.}
A systematic model grid for different types of M-dwarf stars is currently 
under production. 
Here, we present exemplary models with $T_\mathrm{eff} = 3240$\,K and 
$\log g = 4.5$, which represent (weak-field) analogues of AD Leo. 
The models have a horizontal extent of 1950\,km\,$\times$\,1950\,km and 
reach from -700\,km in the convection zone to +1000\,km in the chromosphere. 
All models start from a hydrodynamic simulation snapshot that has been relaxed 
from its initial conditions. 
The initial magnetic field, which is superimposed on the hydrodynamic model, 
is either unipolar and vertical or consists of regions of vertical field with 
opposite polarity (2\,$\times$\,2 checkerboard). 
The individual models have initial magnetic field strengths ranging 
from 10\,G to 500\,G.

\begin{figure}
\includegraphics[width=7.5cm]{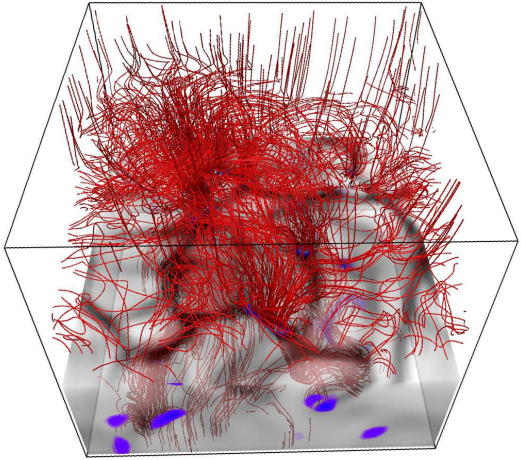}
\caption{Visualisation of a 3-D model snapshot for a M-dwarf model with 
$T_\mathrm{eff} = 3240$\,K, $\log g = 4.5$, and an initial magnetic 
field strength of $|B|_0 = 100$\,G ($2\,\times\,2$ checkerboard). 
The bottom of the photosphere and the convection zone are grey-shaded, 
whereas regions of high magnetic field strength are coloured bluish. 
The red lines represent magnetic field lines. 
This image was produced with VAPOR 
(Clyne et al. 2007; Clyne \& Rast 2005). 
}
\label{fig:vaporfield}
\end{figure}
\section{Atmospheric structure and dynamics} 

Under the conditions in the low photosphere outside strong magnetic flux concentrations 
(i.e., in quiet regions), the magnetic field is effectively "frozen-in".  
Consequently, the flow field generated by the surface convection
continuously rearranges the magnetic field -- like on the Sun.  
The initial magnetic field is swept into the intergranular lanes, where it is 
concentrated in form of continuous sheets and knots.  
Especially in lane vertices, this process produces magnetic flux concentrations with 
field strengths in excess of 1\,kG.  
%
The initial field is thus quickly transformed into a complicated non-uniform 
field.
The complex geometry cannot sufficiently be described as a set of idealized "flux tubes".

The magnetic field expands from these photospheric footpoints into the chromosphere, 
where it funnels out and fills the whole volume. 
The chromopheric magnetic field nevertheless exhibits 
a fine-structure that varies on short time scales. 
In the mixed polarity runs, magnetic loops form which connect footpoint areas of 
opposite polarity. 
The resulting 3-D structure consists of entangled field lines which are often twisted and 
rapidly change in time (see Fig.~\ref{fig:vaporfield}). 
As a consequence, the  model chromosphere varies strongly in space and time, resulting in 
strong variations in the physical state of the atmospheric plasma.
Like for the Sun, a broad distribution of gas temperatures is found in the 
upper layers. 
While upwards propagating shock fronts produce high temperature peaks in the 
chromosphere, the plasma in the post-shock regions cools to low temperatures. 
This non-uniform temperature distribution has to be taken into account when 
calculating meaningful average atmospheric stratifications.

\section{Synthetic intensity maps} 

We illustrate the consequences of the complex atmospheric structure for the 
interpretation of observations. 
To this aim, we apply radiative transfer codes to three model snapshots for 
generating synthetic intensity maps, which can be interpreted as virtual 
observations of the model.

\begin{figure}
\includegraphics[width=\columnwidth]{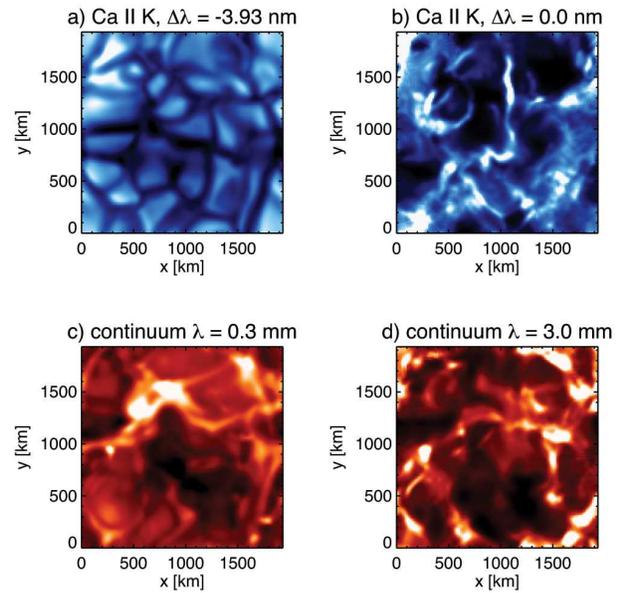}
\caption{Synthetic intensity images for a selected model snapshot. 
\textit{Top:} Ca\,II\,K line wing (left) and line core (right).  
\textit{Bottom:}  continua at 0.3\,mm (left) and 3.0\,mm (right).}
\label{fig:intensitymaps}
\end{figure}
\begin{figure}
\begin{center}
\includegraphics[width=7.5cm]{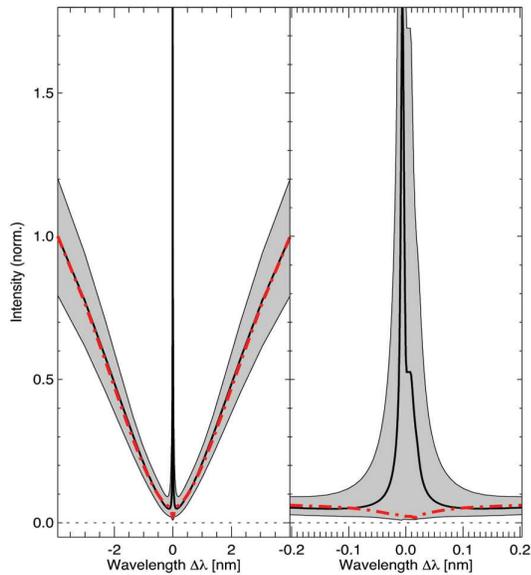}
\end{center}
\caption{Synthetic intensity profiles of the Ca\,II\,K spectral line for a 
selected 3-D model snapshot. 
\textit{Left:} The whole line profile. 
\textit{Right:} Close-up of the line core. 
The grey area between the 1 and the 99 percentiles marks the value range 
covered by the majority of the individual locations in the model. 
The corresponding 3-D average (thick black line, averaged horizontally over all 
columns in the 3-D model) exhibits a central emission peak. 
In contrast, the horizontally averaged model, i.e. the corresponding 1-D 
mean atmosphere, results in a spectral line profile with no emission peak 
(red dot-dashed line). 
}
\label{fig:caprof}
\end{figure}

\paragraph{Calcium.}
First, we use MULTI\_3D 
(Carlsson 1986),
which provides the detailed solution of the radiative transfer equation 
in non-LTE (non-local thermodynamic equilibrium). 
The output contains intensity cubes ($I = f (x, y, \lambda)$) for 5 spectral lines 
(Ca\,II\,H, K, IRT) for each selected snapshot. 
Exemplary intensity maps in the  line wing and in the line core are 
shown in Fig.~\ref{fig:intensitymaps}. 
The line wing image essentially shows the granulation pattern originating from the 
low photosphere. 
The scene is completely different in the line core image. 
The interplay of magnetic fields and propagating shock waves causes an intermittent 
pattern of bright filamentary threads and dark regions, which mostly 
correspond to hot shock fronts and cool post-shock regions, respectively. 
The spectral line profiles accordingly vary significantly for the different 
locations. 
The intensity range covered by the individual Ca\,II\,K lines profiles is illustrated in 
Fig.~\ref{fig:caprof}. 
Despite the large spatial variations, the spatially averaged line profile still 
exhibits a strong central emission reversal peak. 
For comparison, the model snapshot has been averaged horizontally, resulting in 
a one-dimensional model atmosphere. 
When computing the Ca\,II\,K line profile from this 1-D model atmosphere, again 
using MULTI\_3D, it does not show a central emission peak, which is in strong 
contrast to the full 3-D result.
Similar 1-D average atmospheres for other model snapshots can show emission in the 
line core but the differences to the 3-D result are always significant. 
It demonstrates that spatial averaging removes the strong fluctuations connected 
to the dynamic small-scale structure in the model chromosphere, resulting in 
synthetic spectral line profiles that carry only limited if not misleading 
information.

\paragraph{(Sub-)millimeter continua.}
The continuum intensity at (sub-)millimeter wavelengths is 
another promising way to map the chromospheric layers.   
The large diagnostic potential has been demonstrated for the Sun 
(Wedemeyer-B\"ohm et al. 2007).
We use the radiative transfer code LINFOR3D (M.~Steffen, 
 \textit{{http://www.aip.de/$\sim$mst/linfor3D$\_$main.html})} 
to calculate the continuum intensity for wavelengths between 
0.3\,mm and 9.0\,mm, which lie in the range observable with the 
Atacama Large Millimeter Array (ALMA).  
Two exemplary intensity maps are shown in Fig.~\ref{fig:intensitymaps}. 
The average height in the atmosphere, where the intensity is formed,      
increases with increasing wavelength. 
For a plane-parallel stratified atmosphere, it would essentially map 
different chromospheric layers at the  different wavelengths. 
However, the intermittent atmospheric structure results in large 
differences in the effective formation height range. 
At the same wavelength, it is possible to see deeper into the 
atmosphere at certain locations than at others. 
The effective formation height range thus varies spatially and temporally
and may make it difficult to derive a meaningful atmospheric 
stratification from  disk-averaged 
continuum intensity observations.

\begin{figure}
\begin{center}
\includegraphics[width=\columnwidth]{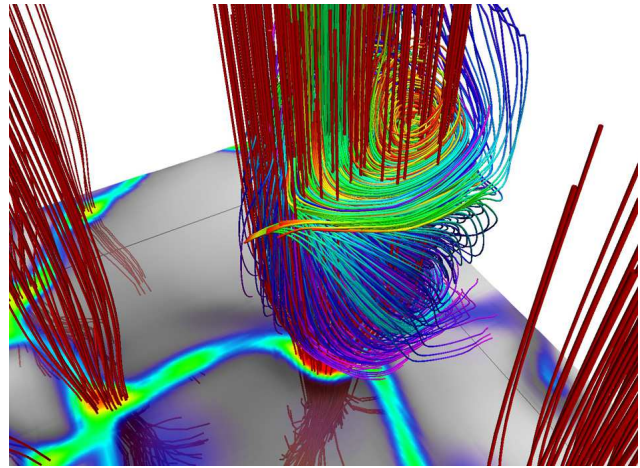}
\end{center}
\caption{Visualisation of a 3-D model snapshot for a M-dwarf model with 
$T_\mathrm{eff} = 3240$\,K, $\log g = 4.5$, and an initial vertical magnetic 
field strength of $|B|_0 = 100$\,G ($2\,\times\,2$ checkerboard). 
The absolute magnetic field strength is color-coded in a plane that represents 
the bottom of the photosphere.
The magnetic field lines (red) are concentrated in the intergranular lanes 
and funnel out in the atmosphere above. 
The plasma in the chromosphere follows the rotating magnetic fields structures, 
resulting in spiral trajectories (green-blue lines). 
It is the first time that such a ``magnetic tornado'' has been found in a 
model of a cool star. 
}
\label{fig:tornado}
\end{figure}

\section{Implications for one-dimensional models} 

Fuhrmeister et al. (2005) 
provide a semi-empirical model atmosphere for AD~Leo, 
in which the gas temperature first decreases with height in the photosphere 
and increases essentially linearly in the chromosphere, followed by a characteristic 
jump in the transition region. 
The range of temperatures found in our 3-D model photosphere and with it the 
corresponding average match the stratification by 
Fuhrmeister et al. (2005)
quite well since the temperature fluctuations are only moderate in that layer. 
The temperature range drastically broadens in the 3-D model chromosphere, owing to 
the onset of shock formation. 
Peak temperatures on the order of 5500\,K are found in many locations throughout 
the chromosphere, resulting in a height-dependent distribution that overlaps with 
the semi-empirical model. 
The arithmetic horizontally averaged temperature, however, increases only mildly 
with height because hot and cold regions cancel.  
The temperature stratification obviously depends on how the temperatures enter the 
average. 
The relation between (local) gas temperature and emergent intensity can be 
strongly non-linear for many spectral lines that are formed in the chromosphere, 
incl. the lines of Ca\,II. 
Deriving a one-dimensional temperature stratification from spatially unresolved 
spectra can thus effectively lead to a higher weighting of the high temperatures 
connected to shock fronts. 

Additional complications include largely varying formation height ranges, which 
makes it difficult to assign a definite height to an empirically derived 
temperature value. 
Consequently, the question arises how meaningful such 1-D average stratifications 
are.

\section{Small-scale dynamics} 
Numerical simulations serve as a tool for analyzing and predicting physical processes. 
The Sun plays an important role as a reference. 
The applied numerical methods can be tested by detailed comparisons of 
solar models with spatially resolved observations of the Sun. 
For cool stars, adjustments and additional physics may be needed but nevertheless
there are many qualitative similarities between solar and stellar models. 
A recent example are so-called magnetic tornadoes on the Sun
(Wedemeyer-B{\"o}hm et al. 2012). 
They are generated by the interplay of magnetic fields and vortex flows in the 
photosphere of the Sun. 
These vortex flows develop due to the conservation of angular momentum of 
cooled plasma in the downdrafts near the surface.
In magnetically quiet Sun regions, the magnetic field is advected with the 
photospheric flows, which forces the magnetic footpoints to rotate. 
The magnetic field mediates the rotation into the upper layers, where the 
situation reverses and the plasma follows the rotating field, resulting in 
spiral trajectories. 
The net energy transport associated with these tornado-like events may 
significantly contribute to the heating of the solar corona.

The models presented here suggest that photospheric vortex 
flows are abundant on the surface of cool M-dwarfs, too, like it was 
already found by 
Ludwig et al. (2002, 2006).
As on the Sun, they force the magnetic footpoints to rotate and thus generate 
magnetic tornadoes in the chromosphere above. 
An example is illustrated in Fig.~\ref{fig:tornado}. 
It is plausible that these small-scale tornadoes are common on M-dwarf
stars, too, which could be of importance for the energy balance of 
their outer atmospheric layers.

\section{Conclusions}
Already the first model generation clearly demonstrates that M-dwarf atmospheres 
are very dynamic and highly structured on small spatial and temporal scales.  
Consequently, a detailed 3-D treatment is required for a realistic description. 
One-dimensional model atmospheres, which are constructed from a comparison to 
observed spectra, reproduce certain features but, by nature, cannot account for 
important properties of the atmospheres. 
Nevertheless, detailed 1-D models will remain complementary to 3-D models until 
all required physical processes can be treated numerically in sufficient detail. 
Such advanced 3-D models will be essential for understanding the magnetic activity 
of M-type dwarf stars. 
 

\acknowledgements 
This work was supported with a grant of the Research Council of Norway 
(No.~208011/F50, \textit{``Magnetic Activity of the Atmospheres of M-type Dwarf Stars''}).

%
\def\apj{ApJ}%
\def\apjl{ApJ}
\def\aap{A\&A}
\def\ssr{Space~Sci.~Rev.}
\def\nat{Nature}
\bibliographystyle{an}

%

\end{document}